\documentclass[dvipdfmx]{PoS}

\title{
  Lattice QCD study of partial restoration of chiral symmetry in the flux-tube 
}  

\ShortTitle{
  Lattice QCD study of partial restoration of chiral symmetry in the flux-tube 
}

\author{\speaker{T.~Iritani}$^a$, G.~Cossu$^a$, and S.~Hashimoto$^{ab}$ \\
\llap{$^a$}High Energy Accelerator Research Organization (KEK),
  Ibaraki 305-0801, Japan \\
  \llap{$^b$}School of High Energy Accelerator Science,
  The Graduate University for Advanced Studies (Sokendai),
  Ibaraki 305-0801, Japan \\
  E-mail: \email{iritani@post.kek.jp}
}

\abstract{
  Using the overlap-Dirac eigenmodes,
  we study the spatial distribution of the chiral condensate
  around static color sources in lattice QCD.
  Between the color sources, there appears a color-flux tube,
  which leads a linear confining potential.
  By measuring a local value of the chiral condensate,
  we show that the magnitude of the condensate
  is reduced inside the flux-tube
  for both quark-antiquark and three-quark systems.
  These results suggest that 
  chiral symmetry is partially restored in the flux-tube.
  The reduction of the condensate is estimated
  to be about 20 $\sim$ 30\% at the center of the flux.
}

\FullConference{XV International Conference on Hadron Spectroscopy-Hadron 2013\\
		4-8 November 2013\\
		Nara, Japan }

\begin{document}

\section{Introduction}

Spontaneous chiral symmetry breaking
is one of the most important properties of QCD.
An order parameter of the symmetry
is the chiral condensate $\langle \bar{q}q \rangle$,
which has a large negative value in non-perturbative vacuum,
while it would be modified at finite temperature or density.

In addition to chiral symmetry breaking,
quark confinement is also an important nature of the QCD vacuum.
Confinement is characterized by linear interquark potential,
which is originated from a flux-tube formation between quarks.
This structure can be observed 
by the action/energy densities around static color sources
in lattice QCD simulation
\cite{Bali:1995,Haymaker:1996,Bissey:2007,Yamamoto:2009}.

Regarding a possible link between chiral symmetry breaking and confinement,
it is interesting to discuss the chiral symmetry 
in the confined system.
Due to color electric and magnetic fields inside the color flux,
non-perturbative properties of QCD could be modified inside.

In this paper, we investigate the chiral condensate
around the static color sources
using the eigenfunctions of the overlap-Dirac fermion operator
in lattice QCD \cite{Iritani:2013rla}.
We discuss the change of chiral symmetry breaking inside the color flux
for both quark-antiquark and three-quark systems.

\section{Partial restoration of chiral symmetry inside the flux-tube}

\subsection{Local chiral condensate}
In order to discuss chiral symmetry inside the color flux,
we define a local chiral condensate 
$\bar{q}q(x)$, which is expressed by means of the Dirac eigenmodes.
In this paper, we use the overlap-Dirac eigenmodes
on 2+1-flavor dynamical overlap-fermion configurations
generated by the JLQCD Collaboration \cite{JLQCD}.
The massless overlap-Dirac operator is given by
\begin{equation}
D_{\rm ov}(0) = m_0 \left[ 1 + \gamma_5 \mathrm{sgn} \ H_W(-m_0) \right],
\label{eq:overlap-Dirac-operator}
\end{equation}
using the hermitian Wilson-Dirac operator $H_W(-m_0) = \gamma_5 D_W(-m_0)$ 
and a sign function \cite{Neuberger:1998}.
This formalism keeps the exact chiral symmetry on the lattice
\cite{GinspargWilson}.

Using the overlap-Dirac eigenfunction $\psi_\lambda(x)$,
that satisfies ${D}_{\rm ov}(0) \psi_\lambda = \lambda \psi_\lambda$,
a local chiral condensate is defined as
\begin{equation}
\bar{q}q(x) = - \sum_\lambda \frac{\psi_\lambda^\dagger(x)\psi_\lambda(x)}{m_q
  + (1 + \frac{m_q}{2m_0})\lambda},
\label{eq:localChiralCondensate}
\end{equation}
with eigenvalues $\lambda$ and a quark mass $m_q$.
The chiral condensate $\langle \bar{q}q \rangle$ is given by 
an expectation value of $\bar{q}q(x)$.
On a given gauge configuration in the broken phase,
the spatial distribution of $\bar{q}q(x)$ show local clusters,
which correspond to the location of the (anti-)instantons
\cite{Schafer:1996wv}.

\subsection{Chiral condensate in quark-antiquark system}

First, we discuss quark-antiquark ($\mathrm{Q\bar{Q}}$) system.
Similar to the flux-tube measurement in lattice QCD
\cite{Bali:1995,Haymaker:1996,Bissey:2007,Yamamoto:2009},
we analyze the spatial distribution of the chiral condensate $\bar{q}q(x)$ around
two static color sources as
\begin{equation}
    \langle \bar{q}q(x) \rangle_W 
    \equiv \frac{\langle\bar{q}q(x) W(R,T)\rangle}{\langle W(R,T)\rangle}
    - \langle \bar{q}q \rangle,
  \label{eq:diffLocalChiral}
\end{equation}
where $W(R,T)$ is the Wilson loop with the size of $R\times T$.
Here, we use a low-mode truncated condensate 
$\bar{q}q^{(N)}(x)$ by truncating the sum in Eq.~(\ref{eq:localChiralCondensate}).

In this paper, we use $24^3 \times 48$ lattice at $\beta = 2.3$,
which corresponds to a lattice spacing $a^{-1} = 1.759(10)$~GeV.
The dynamical quark masses are $m_{ud} = 0.015$ and $m_s = 0.080$,
and the global topological charge is fixed at $Q = 0$.

\begin{figure}
  \centering
  \includegraphics[width=0.49\textwidth,clip]{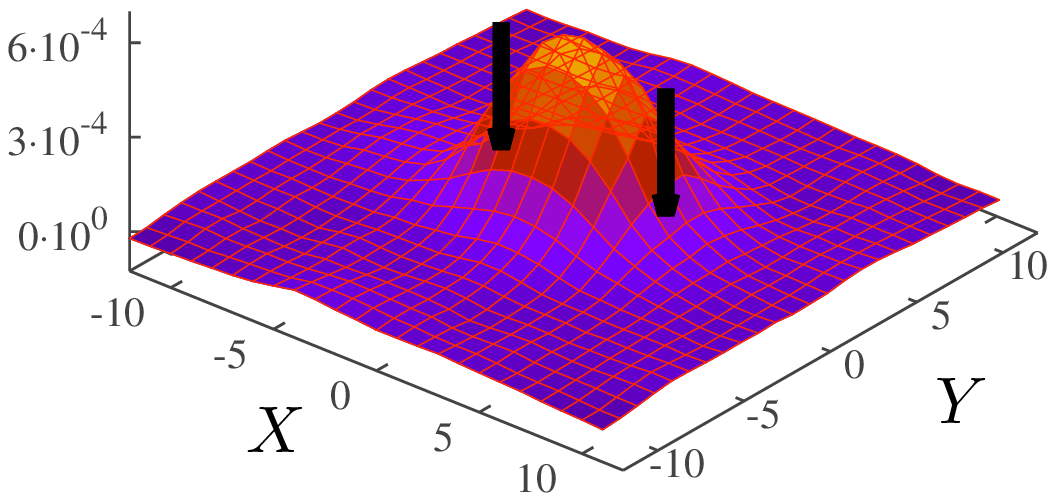}
  \includegraphics[width=0.49\textwidth,clip]{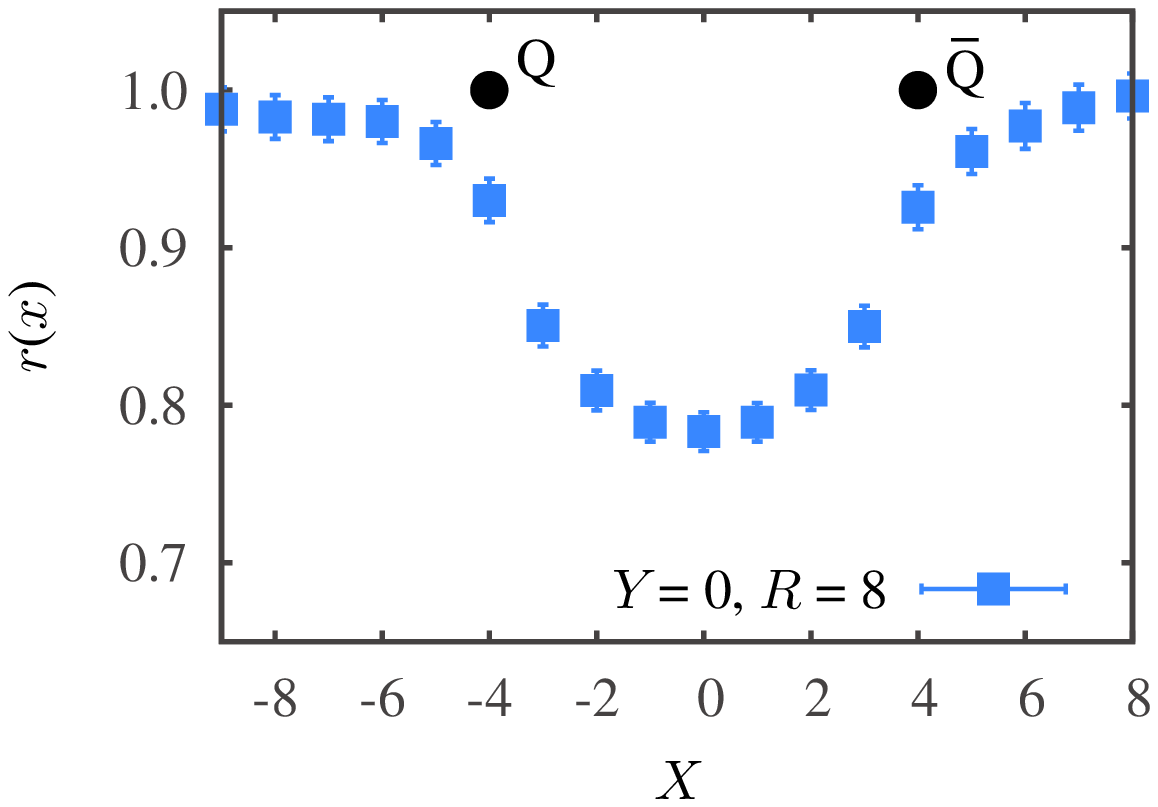}
  \caption{
    \label{fig:localChiral}
    (a) The spatial distribution of the local chiral condensate 
    $\langle \bar{q}q(x) \rangle_{W}$ around the Wilson loop $W(R,T)$ with $R = 8$.
    The vertical bars denote the position of the color sources
    at $(4,0)$ and $(-4,0)$.
    (b) The cross-section of the ratio $r(x)$ along the tube ($Y = 0$),
        the circles denote the position of the color sources.
      }
\end{figure}

Figure \ref{fig:localChiral} (a)
shows $\langle \bar{q}q(x) \rangle_W^{(N)}$ at a separation $R = 8$ 
with a number of eigenmodes $N = 160$, 
the color sources are located at $(4, 0)$ and $(-4, 0)$ on $XY$-plane,
and the valence quark mass $m_q = 0.015$ \cite{JLQCD}. 
In order to improve the signal of the Wilson loop, 
we use the APE smearing for the spatial link-variables.
$W(R,T)$ is measured at $T = 4$, and the number of configuration is 50.

As can be seen in Fig.~\ref{fig:localChiral} (a),
the change of the chiral condensate is positive
between the color sources. 
It implies that the magnitude of the condensate is reduced
inside the flux-tube,
since $\langle \bar{q}q \rangle$ is negative in QCD vacuum.

Since $\bar{q}q$ is a divergent operator,
we need to renormalize for a quantitative analysis.
Considering the mode truncation as a regularization method \cite{Noaki:2009xi},
the power divergence is parametrized as
\begin{equation}
  \langle \bar{q}q \rangle^{(N)} = 
  \langle \bar{q}q^{\rm (subt)} \rangle  
  + {c_1^{(N)}m_q/a^2} + c_2^{(N)}m_q^3,
  \label{eq:subtractChiral}
\end{equation}
due to the exact chiral symmetry of the overlap-Dirac operator.
The subtracted condensate $\langle \bar{q}q^{\rm (subt)} \rangle$ 
becomes finite up to logarithmic divergence,
which can be obtained by fitting $\langle \bar{q}q \rangle^{(N)}$ as a function of $m_q$
\cite{Noaki:2009xi}.
The remaining divergence also cancels taking the ratio
\begin{equation}
  r(x) \equiv \frac{\langle \bar{q}q^{\rm (subt)}(x)W(R,T) \rangle}
  {\langle \bar{q}q^{\rm (subt)} \rangle \langle W(R,T)\rangle}.
  \label{eq:reductionRatio}
\end{equation}

Figure \ref{fig:localChiral} (b) shows the cross-section of 
the ratio $r(x)$ along the tube.
The ratio $r(x)$ clearly shows that chiral symmetry is partially restored 
between the color sources.
The reduction is about 20\% at around the center of 
quark-antiquark system with a separation $R \simeq 0.9$~fm.

\subsection{Chiral condensate in three-quark system}

\begin{figure}
  \centering
      \includegraphics[width=0.49\textwidth,clip]{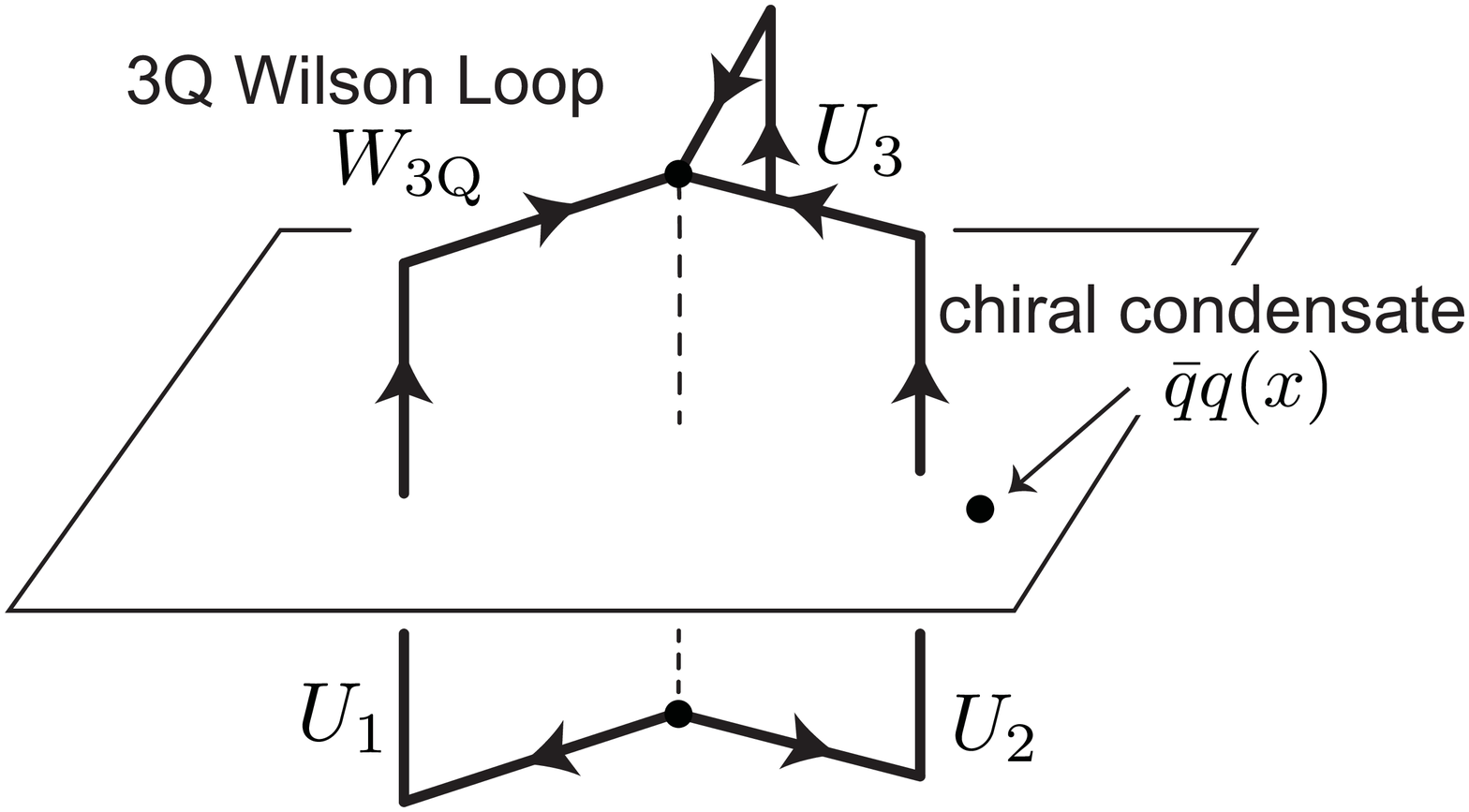}
      \includegraphics[width=0.49\textwidth,clip]{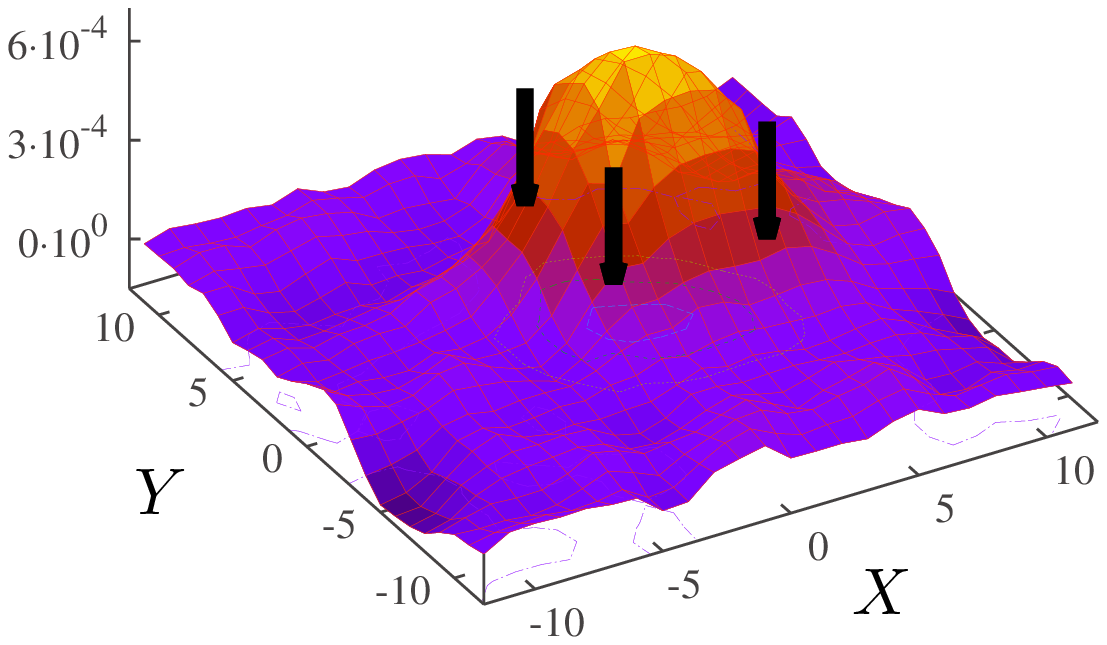}
  \caption{
    \label{fig:3Qflux}
    (a) The schematic picture of the spatial distribution measurement 
      of the chiral condensate $\bar{q}q(x)$ in the 3Q system.
    (b) The local chiral condensate 
    $\langle \bar{q}q(x) \rangle_{W_{\rm 3Q}}$ around the 3Q-Wilson loop $W_{\rm 3Q}$.
    The color sources are located at $(X,Y) = (0,0), (6,0)$, and $(0,6)$,
  which are denoted by the vertical bars.}
\end{figure}

Next, we investigate the three-quark (3Q) system.
The 3Q-Wilson loop $W_{\rm 3Q}$ is defined by
\begin{equation}
  W_{\rm 3Q} \equiv \frac{1}{3}\varepsilon_{abc} \varepsilon_{a'b'c'}
  U_1^{aa'}U_2^{bb'}U_3^{cc'},
  \label{}
\end{equation}
with the path-ordered product $U_k \equiv \int_{\Gamma_k}e^{iagA_k} dx$
along the path $\Gamma_k$ \cite{Takahashi:2003}. 
In a similar manner as the $\mathrm{Q\bar{Q}}$-system, 
we measure the chiral condensate around the 3Q-Wilson loop 
$\langle \bar{q}q(x) \rangle_{W_{\rm 3Q}}$ 
by replacing $W(R,T)$ with $W_{\rm 3Q}$ in Eq.~(\ref{eq:diffLocalChiral}).
The schematic picture of the measurement is shown in Fig.~\ref{fig:3Qflux} (a).

Figure \ref{fig:3Qflux} (b) shows  $\langle \bar{q}q(x) \rangle_{W_{\rm 3Q}}$
with color sources at $(0,0)$, $(6,0)$, and $(0,6)$ on the $XY$-plane.
$\langle \bar{q}q(x) \rangle_{W_{\rm 3Q}}$
becomes positive among the color sources, 
which indicates the partial restoration 
of chiral symmetry like the $\mathrm{Q\bar{Q}}$-system case.

\begin{figure}
  \centering
      \includegraphics[width=0.49\textwidth,clip]{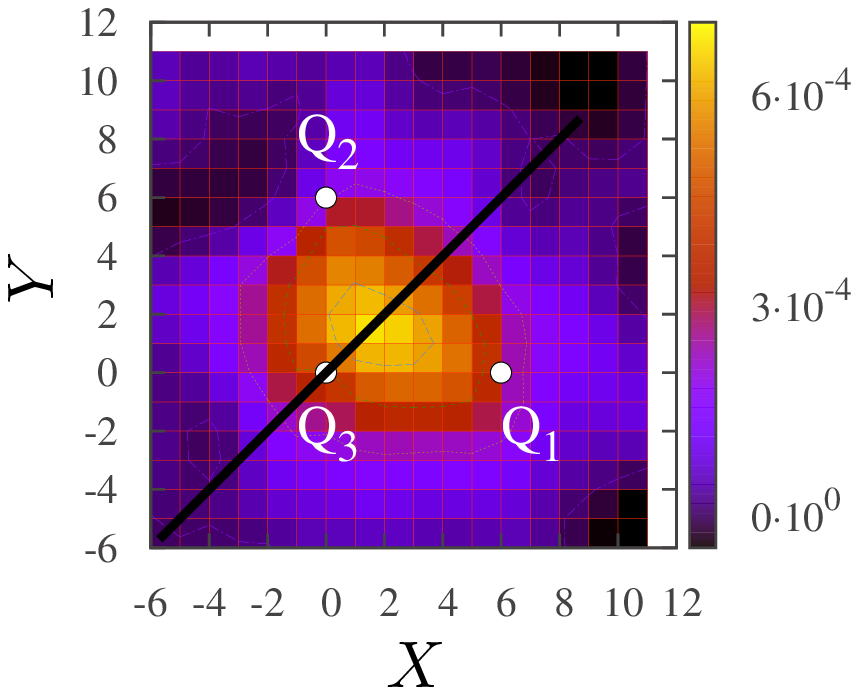}
      \includegraphics[width=0.49\textwidth,clip]{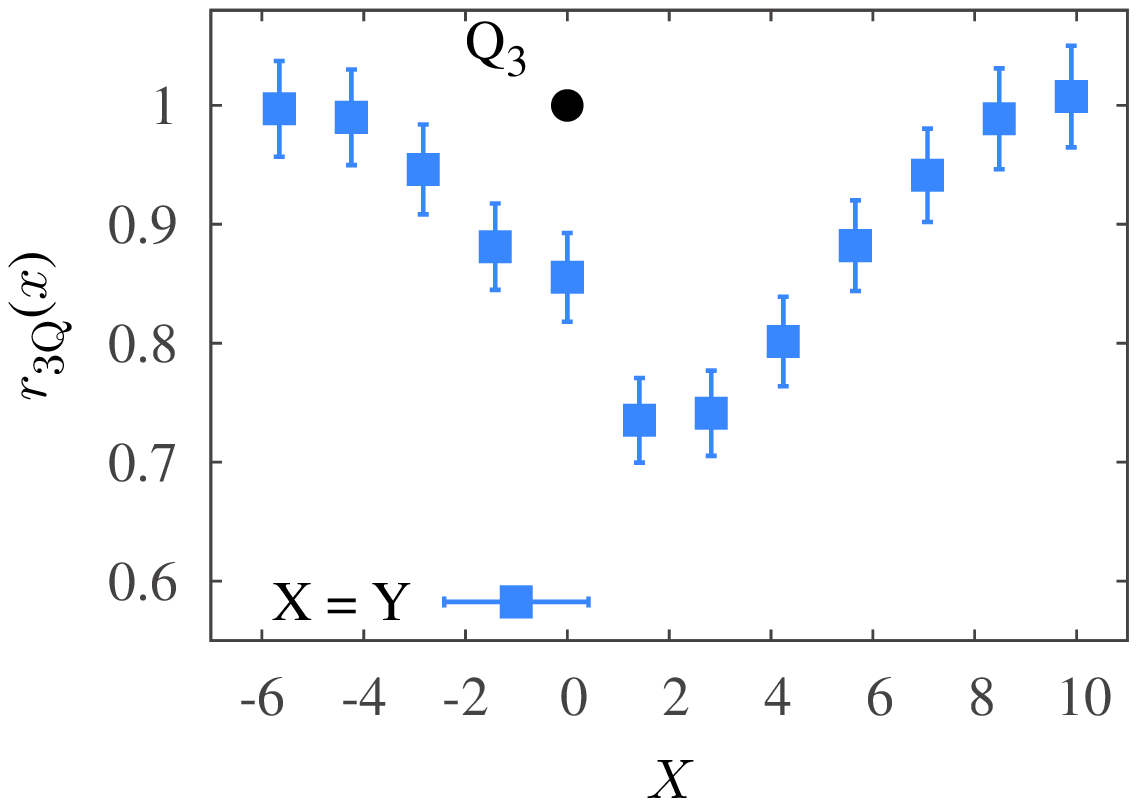}
  \caption{
    \label{fig:ratio3Q}
    (a) 
    A heat map of the chiral condensate around three color sources,
    a black line denotes a slice shown in right figure.
    (b) 
    The cross-section of the ratio $r(x)$ along the line $X = Y$.
    One of the color sources is located at the origin $(X,Y) = (0,0)$.
  }
\end{figure}

Finally, we estimate the ratio of chiral condensate $r_{\rm 3Q}(x)$
by replacing $W(R,T)$ with $W_{\rm 3Q}$ in Eq.~(\ref{eq:reductionRatio}).
Considering a cross-section along a line which is depicted in Fig.~\ref{fig:ratio3Q} (a),
we plot $r_{\rm 3Q}(x)$ in Fig.~\ref{fig:ratio3Q} (b).
In this case,
the reduction of chiral condensate is estimated about 30\% at the center of the 3Q-flux.

\section{Summary and Discussion}

In this work, we investigate chiral symmetry breaking
around the static color sources in lattice QCD.
Using overlap-Dirac eigenmodes,
we study the local chiral condensate around the Wilson and the 3Q-Wilson loops.
Among the static color sources,
there appears a color-flux tube, which produces a linear confining potential.
By measuring the spatial distribution of the chiral condensate,
we find a reduction of the chiral condensate inside the flux 
for both quark-antiquark and three-quark systems.
These results may be interpreted as 
partial restoration of chiral symmetry inside ``hadrons.''
Similar results are also reported 
by measuring the chiral condensate around the Polyakov loop in lattice QCD
\cite{Faber:1993}, and 
the flux-tube effects using
Nambu-Jona-Lasinio model \cite{Suganuma:1990nn}.

\section*{Acknowledgements}
  The lattice QCD calculations have been done on SR16000 at 
  High Energy Accelerator Research Organization (KEK)
  under a support of its Large Scale Simulation Program (No. 12-05).
  This work is supported in part by the Grant-in-Aid of the Japanese
  Ministry of Education (No. 21674002),
  and the SPIRE (Strategic Program for Innovative REsearch) Field 5 project.


\begin{thebibliography}{99}

  \bibitem{Bali:1995}
    G.~S.~Bali, K.~Schilling and C.~Schlichter,
    \emph{Phys.\ Rev.} {\bf D51} (1995) 5165 [hep-lat/9409005].

  \bibitem{Haymaker:1996}
    R.~W.~Haymaker, V.~Singh, Y.~-C.~Peng and J.~Wosiek,
    \emph{Phys.\ Rev.} {\bf D53} (1996) 389 [hep-lat/9406021].

  \bibitem{Bissey:2007}
    F.~Bissey, F-G.~Cao, A.~R.~Kitson, A.~I.~Signal,
    D.~B.~Leinweber, B.~G.~Lasscock and A.~G.~Williams,
    \emph{Phys.\ Rev.} {\bf D76} (2007) 114512, [hep-lat/0606016].

  \bibitem{Yamamoto:2009}
    A.~Yamamoto,
    \emph{Phys.\ Lett.\ B} {\bf 688} (2010) 345 [arXiv:0906.2618 [hep-lat]].

  \bibitem{Iritani:2013rla}
    T.~Iritani, G.~Cossu and S.~Hashimoto,
    \pos{PoS (Lattice 2013) 376} [arXiv:1311.0218 [hep-lat]].

  \bibitem{JLQCD}
    S.~Aoki et al. (JLQCD and TWQCD Collaborations), 
    \emph{Prog. Theor. Exp. Phys.} {\bf 2012} (2012) 01A106.

  \bibitem{Neuberger:1998}
    H.~Neuberger, \emph{Phys. Lett. B} {\bf 417} (1998) 141
    [hep-lat/9707022];
    \emph{ibid.} \textbf{427} (1998) 353 [hep-lat/9801031].

  \bibitem{GinspargWilson}
    P.~H.~Ginsparg and K.~G.~Wilson,
    \emph{Phys. Rev.} {\bf D25} (1982) 2649.

  \bibitem{Schafer:1996wv}
    T.~Sch\"afer and E.~V.~Shuryak,
    \emph{Rev.\ Mod.\ Phys.} {\bf 70} (1998) 323 [hep-ph/9610451].

  \bibitem{Noaki:2009xi} 
    J.~Noaki, T.~W.~Chiu, H.~Fukaya, S.~Hashimoto, H.~Matsufuru, 
    T.~Onogi, E.~Shintani and N.~Yamada,
    \emph{Phys.\ Rev.} {\bf D81} (2010) 034502
    [arXiv:0907.2751 [hep-lat]].

  \bibitem{Takahashi:2003} 
    T.~T.~Takahashi and H.~Suganuma, 
    \emph{Phys. Rev. Lett.} {\bf 86} (2001) 18;
    \emph{Phys. Rev.} {\bf D65} (2002) 114509.

  \bibitem{Faber:1993}
    M.~Faber, M.~Schaler, and H.~Gausterer,
    \emph{Phys. Lett. B} {\bf 317} (1993) 409;
    W.~Sakuler, W.~Burger, M.~Faber, H.~Markum, M.~Muller, P.~De Forcrand,
    A.~Nakamura and I.~O.~Stamatescu,
    \emph{Phys.\ Lett.\ B} {\bf 276} (1992) 155;
    S.~Thurner, M.~Feurstein, H.~Markum, and W.~Sakuler,
    \emph{Phys.\ Rev.} {\bf D54} (1996) 3457;
    K.~H\"ubner, 
    \pos{PoS (Lattice 2007) 193} [arXiv:0709.1467 [hep-lat]].

  \bibitem{Suganuma:1990nn} 
    H.~Suganuma and T.~Tatsumi,
    \emph{Annals Phys.}\  {\bf 208} (1991) 470;
    \emph{Phys.\ Lett.\ B} {\bf 269} (1991) 371;
    \emph{Prog. Theor. Phys.} {\bf 90} (1993) 379.

\end{thebibliography}
\end{document}